\documentclass[preprint,aps,prd,showpacs,groupedaddress,superscriptaddress,%
floatfix,nofootinbib]{revtex4}%
\usepackage{graphicx}
\usepackage{amsmath}
\usepackage{bm}
\usepackage{epsfig}
\begin{document}
\title{Optimal spin-quantization axes \\
	for quarkonium with large transverse momentum 
}
\author{Eric~Braaten}
\affiliation{Physics Department, Ohio State University, 
Columbus, Ohio 43210, USA}
\author{Daekyoung~Kang}
\affiliation{Physics Department, Ohio State University, 
Columbus, Ohio 43210, USA}
\author{Jungil~Lee}
\affiliation{Department of Physics, Korea University, Seoul 136-701, Korea}
\author{Chaehyun~Yu}
\affiliation{Department of Physics, Korea University, Seoul 136-701, Korea}

\begin{abstract}
The gluon collision process that creates
a heavy-quark-antiquark pair with small relative momentum 
and large transverse momentum predicts at leading-order 
in the QCD coupling constant that the transverse polarization 
of the pair should increase with its transverse momentum. 
Measurements at the Fermilab Tevatron of the polarization of 
charmonium and bottomonium states with respect to a particular 
spin-quantization axis are inconsistent with this prediction.
However the predicted rate of approach to complete 
transverse polarization depends on the choice of spin-quantization axis.
We introduce axes that maximize and minimize the transverse 
polarization from the leading-order gluon collision process. 
They are determined by the direction of the jet that provides 
most of the balancing transverse momentum.
\end{abstract}

\maketitle

Important information about the fundamental interactions 
between elementary particles is encoded in the spins of particles 
produced in high-energy collisions.  It is difficult to access 
that information, because the spin cannot be measured directly.
The measurement of the spin state of a stable particle is usually
not possible in a high-energy collider experiment. 
Information about the spin state of an unstable particle can 
be inferred from the angular distribution of its decay products.
That angular distribution can be strongly correlated with the 
momenta of other particles in the final state, and if they are not 
measured, information about the spin is diluted.
Hadron collisions have the additional complication that the
fundamental interactions involve collisions of partons with 
longitudinal momenta that must be integrated over, which
further dilutes the information about the spin.  

The accessible information about the spin
of the unstable particle could in principle be obtained by 
measuring the complete angular distribution of its decay products
for all possible values of the other relevant final-state
variables.  In practice, there is usually not enough data to 
make this feasible.  The challenge is to find simple 
measurements that extract as much information as possible
about the spin.  In a two-body decay, the simplest 
measurement is the distribution of a polar angle $\theta$ 
with respect to a specified axis.  If that axis is used 
as the {\it spin-quantization axis} (SQA) for the unstable particle,
the $\theta$ distribution gives the probabilities for the 
projection of the spin along that axis.
Additional information about the spin could be obtained by also
measuring the polar angle distribution about a second SQA.
These SQA's can be chosen as functions of other final-state 
variables to maximize the information about the spin.

An interesting unsolved problem in high-energy physics 
is the spin dependence of the production rates for heavy 
quarkonia, whose constituents are a heavy (charm or bottom) 
quark and its antiquark.  The nonrelativistic QCD (NRQCD) 
factorization framework
exploits the large mass of the heavy quark to express a
cross section for inclusive quarkonium production 
as the sum of products of parton cross sections that 
can be calculated using perturbative QCD and 
NRQCD matrix elements that can be measured in other 
experiments \cite{Bodwin:1994jh}.  This approach leads to unambiguous 
predictions for the spin dependence of the cross sections.
Measurements of the spin dependence of charmonium and 
bottomonium states produced at the Tevatron $p \bar p$ collider
\cite{CDF:polznpsi,Acosta:2001gv,D0:2008za} are inconsistent 
with predictions based on leading-order parton cross sections
\cite{Beneke:1996yw,Leibovich:1996pa,Braaten:1999qk,Braaten:2000gw}.
The resolution of the discrepancy may lie in large 
perturbative corrections to the parton cross sections
\cite{GW08,ALM08}.  

Most theoretical predictions and the experimental measurements
of quarkonium polarization have been carried out for a 
single SQA.  
In this paper, we point out that more detailed information
about the spin can be extracted by measuring
the polarization for a pair of SQA's for which the 
predicted polarizations are as different as possible. 
We construct a pair of optimal SQA's for spin-triplet $S$-wave
quarkonium states with large transverse momentum
under the assumption that the dominant parton process is 
a gluon-gluon collision.

To be specific, we consider the production of the 
spin-1 charmonium state $J/\psi$ in hadron collisions.
There is a simple argument that a direct $J/\psi$, 
i.e.\ one that does not come from the decay of a heavier particle,
should be increasingly transversely polarized as its 
transverse momentum $Q_T$ increases \cite{Cho:1994gb}.
Of the leading-order parton processes that 
produce a $c \bar c$ pair with large $Q_T$, 
the gluon collision process 
$g g \to (c \bar c) g$ has the largest cross section.  
At asymptotically large $Q_T$, 
this cross section is dominated by {\it gluon fragmentation},
the production of an almost on-shell gluon by 
$g g \to g g$ followed by the splitting of that virtual gluon 
into a collinear $c \bar c$ pair.  The $c \bar c$ pair is created 
in a color-octet state and is predominantly 
transversely polarized for almost any choice of SQA.
{\it Heavy-quark spin symmetry} implies that the binding of this 
$c \bar c$ pair into a $J/\psi$ is unlikely to change the 
spin states of the heavy quarks.  Thus, the $J/\psi$ 
should be increasingly transversely polarized as $Q_T$ increases.
These qualitative arguments are supported by quantitative 
calculations using the NRQCD factorization formulas with 
leading-order parton cross sections 
\cite{Beneke:1996yw,Leibovich:1996pa,Braaten:1999qk}.
However, measurements by the CDF Collaboration seem to be 
completely incompatible with these predictions \cite{CDF:polznpsi}.

The longitudinal polarization 4-vector $\epsilon_L$ 
for a $J/\psi$  with 4-momentum $Q$ must satisfy
$Q \cdot \epsilon_L = 0$ and $\epsilon_L^2 = -1$.
The most general 4-vector 
satisfying these conditions can be written in the form
\begin{equation}
\epsilon_L^\mu = \tilde X^\mu/\sqrt{-\tilde X^2}, \  \  
\tilde X^\mu = (-g^{\mu \nu} +  Q^\mu Q^\nu/Q^2) X_\nu,
\label{epsilonL-Xtilde}
\end{equation}
where  $X$ is a 4-vector.  The physical interpretation 
of $X$ is that in a $J/\psi$ rest frame,
the direction of the 3-vector $-\bm{X}$ is the SQA
of the $J/\psi$.  If $X$ is timelike, 
the projection of the spin along the SQA 
coincides with the helicity in the rest frame of $X$.
If the $J/\psi$ is produced in the collision
of two hadrons with 4-momenta $P_1$ and $P_2$, 
the SQA is generally chosen 
to lie in the production plane defined by the momenta 
of the colliding hadrons in the $J/\psi$ rest frame. 
Thus, $X$ in Eq.~(\ref{epsilonL-Xtilde}) has the form
\begin{equation}
X^\mu = a P_1^\mu + b P_2^\mu,
\label{X-P1P2}
\end{equation}
where $a$ and $b$ are scalar functions. 
A simple example is the {\it c.m.\ helicity axis} specified by 
$X_{\rm cmh} =  P_1 + P_2$.
The projection of the spin of the $J/\psi$ along this axis 
is its helicity in the center-of-momentum (c.m.) frame 
of the colliding hadrons.  
In the $J/\psi$ rest frame, the SQA specified by 
Eq.~(\ref{X-P1P2}) is antiparallel to the unit vector 
$\bm{\hat X} = 
(a \bm{P}_1 + b \bm{P}_2)/(|a \bm{P}_1 + b \bm{P}_2|)$,
so it is determined by the ratio $a/b$.
If the $J/\psi$ is the only particle in the final state 
whose momentum is measured, then $a/b$ can  
only depend on the 4-vectors $P_1$, $P_2$, and $Q$.
If additional information about the final state is measured, 
$a/b$ can also depend on this information.  

If a cross section is dominated by a specific parton 
process, a natural prescription for a pair of optimal SQA's 
is that they maximize and minimize the transverse cross section 
from that parton subprocess 
at leading order in the QCD coupling constant $\alpha_s$.  
As a simple illustration of a pair of optimal SQA's,
we consider dilepton production by the
Drell-Yan mechanism $q \bar q \to \mu^+ \mu^-$
in the parton model with intrinsic transverse momentum
\cite{Collins:1977iv,Lam:1978pu},
which can be regarded as a model for the effects of soft gluon 
radiation from the colliding partons.
The cross section for a dilepton with longitudinal
polarization vector given by Eqs.~(\ref{epsilonL-Xtilde}) 
and (\ref{X-P1P2}) is 
\begin{equation}
\hat \sigma_L = 
\frac{8 \pi^2 e_q^2 \alpha \langle k_\perp^2 \rangle 
	(a^2 x_2^2 + b^2 x_1^2)}{3 Q^2 (a x_2 - b x_1)^2}
\delta(x_1 x_2 s - Q^2),
\label{sigpart-ave}
\end{equation}
where $x_1$ and $x_2$ are the longitudinal momentum fractions
of the colliding partons, $e_q$ is the electric charge of the quark,
and $\langle k_\perp^2 \rangle$ is the mean-square 
transverse momentum of the colliding partons.
The cross section has been expanded to second order in 
the intrinsic transverse momentum vectors
and then averaged over them.
The longitudinal cross section in 
Eq.~(\ref{sigpart-ave}) depends on the ratio $a/b$.
It is minimized by choosing $a/b = - x_1/x_2$.
It can be maximized by choosing
$a/b = +x_1/x_2$, so that the denominator vanishes.  
The values $a/b = \mp x_1/x_2$ that maximize and minimize the transverse 
cross section depend on the 
parton momentum fractions only through the ratio $x_1/x_2$.
Under the assumption that the cross section
is dominated by $q \bar q \to \mu^+ \mu^-$,
we can derive an expression for $x_1/x_2$ in terms of 
variables that can be directly measured.
At leading order in the intrinsic transverse momentum vectors,
the energy-momentum conservation condition 
$Q = x_1 P_1 + x_2 P_2$  implies 
$x_1/x_2 = Q \cdot P_2/Q \cdot P_1$.
The 4-vectors $X$ in Eq.~(\ref{X-P1P2}) associated 
with the maximal and minimal SQA's can then be expressed as
\begin{equation}
X_{{\rm CS}/\perp {\rm h}}^\mu =  
\frac{P_1^\mu}{Q \cdot P_1} \mp \frac{P_2^\mu}{Q \cdot P_2} .
\label{epsilonL-CS}
\end{equation}
The upper and lower signs correspond to the 
{\it Collins-Soper axis}  \cite{Collins:1977iv} and the
{\it perpendicular helicity axis} 
introduced in Ref.~\cite{Braaten:2008mz}, respectively. 
The perpendicular helicity axis is so named because
the projection of the total spin of the dilepton along 
this SQA coincides with its helicity in the frame obtained 
from the hadron c.m.\ frame by a longitudinal boost that 
makes the dilepton momentum perpendicular to the beam direction.
Lam and Tung pointed out that the Collins-Soper axis 
maximizes the transverse polarization in the parton model 
with intrinsic transverse momentum \cite{Lam:1978pu}. 
That the perpendicular helicity axis minimizes the
transverse polarization in this model does not seem 
to have been pointed out previously.

We now consider the direct production of a $J/\psi$ with 
large transverse momentum $Q_T \gg \Lambda_{\rm QCD}$.
Since the mass $m_c$ of the charm quark is large 
compared to the scale $\Lambda_{\rm QCD}$ of nonperturbative effects 
in QCD, the inclusive cross section can 
be calculated using NRQCD factorization formulas \cite{Bodwin:1994jh}. 
The leading-order parton processes that create a $c \bar c$ pair
with small relative momentum are
$q \bar q \to (c \bar c) g$, $q g \to (c \bar c) q$, 
$\bar q g \to (c \bar c) \bar q$, and 
$g g \to (c \bar c) g$.  The parton process with 
the largest cross section is $g g \to (c \bar c) g$.
This process is further enhanced by the growth of the 
gluon distribution at small values of the parton momentum 
fraction $x$ as the momentum scale increases.
We therefore focus on the parton process $g g \to (c \bar c) g$.
The color-octet $^3S_1$ state of the $c \bar c$ pair 
 $[c\bar{c}_8(^3S_1)]$ dominates 
at asymptotic $Q_T$, because it has a fragmentation contribution.
At leading order in $\alpha_s$, the dependence of the 
longitudinal differential cross section on the scalar functions $a$ 
and $b$ defined by Eq.~(\ref{X-P1P2}) is 
\begin{equation}
Q_0 \frac{d \hat \sigma_L }{d^3Q} \propto 
\frac{a^2 x_2^2 \hat u^2 + b^2 x_1^2 \hat t^2
	+ (a x_2 - b x_1)^2 \hat s^2}
    {(a w_1 + b w_2)^2 - a b Q^2 s},
\label{sig:qqbar2}
\end{equation}
where $w_1 = Q \cdot P_1$, $w_2 = Q \cdot P_2$, $s$ 
is the c.m.\ energy of the colliding hadrons,
and $\hat s = x_1 x_2 s$, $\hat t = 4 m_c^2 - 2 x_1 w_1$, 
and $\hat u = 4 m_c^2 - 2 x_2 w_2$ are the parton Mandelstam variables.
There is a delta function constraint on these variables:
\begin{equation}
2(x_1 w_1 + x_2 w_2) = x_1 x_2 s + 4 m_c^2.
\label{deltacon}
\end{equation}
Minimizing and maximizing Eq.~(\ref{sig:qqbar2}) with respect to $a/b$, 
we get
\begin{eqnarray}
\frac{a}{b}\Big|_{gg,\rm max} &=& 
\frac{x_1(\hat{s}-\hat{t})}{x_2(\hat{s}-\hat{u})},
\label{ab:maxgg}
\\
\frac{a}{b}\Big|_{gg,\rm min} &=& 
\frac{x_1[\hat{s}^2(\hat{t}-\hat{u})+\hat{t}^2(\hat{s}-\hat{u})]}
    {x_2[\hat{s}^2(\hat{t}-\hat{u})+\hat{u}^2(\hat{t}-\hat{s})]}.
\label{ab:mingg}
\end{eqnarray}
At asymptotic transverse momentum $Q_T \gg 2 m_c$,
the corresponding 4-vectors reduce to
\begin{eqnarray}
X^\mu_{gg,\rm max} &\longrightarrow& 
\frac{x_1 P_1^\mu}{x_1 w_1 + 2 x_2 w_2}
+ \frac{x_2 P_2^\mu}{x_2 w_2 + 2 x_1 w_1},
\label{X-maxgg:asympt}
\\
X^\mu_{gg,\rm min} &\longrightarrow& 
\frac{P_1^\mu}{w_1} 
- \frac{P_2^\mu}{w_2} = X_{\rm CS}^\mu.
\label{X-mingg:asympt}
\end{eqnarray}
Note that the minimal $gg$ axis reduces in this limit 
to the Collins-Soper axis.
 
Our optimality criteria were based on the assumption 
that the parton process $g g \to c \bar c_8(^3S_1)+g$ dominates.
If that parton process implies values for $x_1$ and $x_2$
that are determined by a measurable property of the final state, 
we can insert those values into Eqs.~(\ref{ab:maxgg}) and (\ref{ab:mingg})
to obtain optimal SQA's that are experimentally useful. 
The large transverse momentum $Q_T$ of the $c \bar c$ pair is
balanced by that of the recoiling gluon, which 
produces a jet of hadrons with nearly collinear momenta.
The polar angle of the recoiling gluon 
in the hadron c.m.\  frame 
is approximately equal to the polar angle $\theta_{\rm jet}$ of the jet. 
The ratio $x_1/x_2$ can be expressed as a function of 
$\theta_{\rm jet}$ and the
transverse and longitudinal momenta $Q_T$ and $Q_L$ 
of the $J/\psi$ in the hadron c.m.\ frame: 
\begin{equation}
\frac{x_1}{x_2} = 
\frac{(Q_0 + Q_L) \sin \theta_{\rm jet}
	+ Q_T (1 + \cos \theta_{\rm jet})}
    {(Q_0 - Q_L) \sin \theta_{\rm jet}
	+ Q_T (1 - \cos \theta_{\rm jet})} ,
\label{ratio-opt}
\end{equation}
where $Q_0 = (Q_T^2 + Q_L^2 + M_{J/\psi}^2)^{1/2}$.
Given the value of $x_1/x_2$,
the delta function constraint in Eq.~(\ref{deltacon})
can be solved for $x_1$ and $x_2$.
The two-fold ambiguity associated with the root of a quadratic 
polynomial is resolved by choosing the smaller values 
of $x_1$ and $x_2$.
Inserting them into Eqs.~(\ref{ab:maxgg}) and (\ref{ab:mingg}),
we obtain expressions for $a/b$  
that depend on quantities that can be directly measured.
We will refer to the corresponding SQA's as the 
{\it maximal} and {\it minimal $gg$} axes, respectively.
The prescriptions for these optimal SQA's can be extended
beyond leading order in $\alpha_s$ by choosing 
$\theta_{\rm jet}$ in Eq.~(\ref{ratio-opt})
to be the angle in the hadron c.m.\ frame of the jet 
with the largest transverse energy.
Since the direction of the jet
is insensitive to soft gluon radiation and to the 
splitting of a parton into collinear partons,
QCD radiative corrections to the polar angle distributions 
defined by our optimal SQA's can be calculated 
systematically using perturbative QCD.

To illustrate the use of pairs of SQA's,
we consider direct $J/\psi$ production at the Tevatron,
which is a $p \bar p$ collider with c.m.\ energy 1.96~TeV.
We use NRQCD factorization formulas with leading-order parton 
cross sections to calculate the transverse and longitudinal 
cross sections for $J/\psi$.
For the NRQCD matrix elements, we use central CTEQ5L values
from Ref.~\cite{Braaten:1999qk} with $x = 1/2$:
$\langle O_1^{J/\psi}(^3S_1) \rangle = 1.4~{\rm GeV}^3$, 
$\langle O_8^{J/\psi}(^3S_1)\rangle = 0.0039~{\rm GeV}^3$, 
$\langle O_8^{J/\psi}(^1S_0)\rangle = 0.033~{\rm GeV}^3$, and
$\langle O_8^{J/\psi}(^3P_0)\rangle/m_c^2 = 0.0097~{\rm GeV}^3$.
In the parton cross sections, we set $m_c = M_{J/\psi}/2 = 1.5~{\rm GeV}$.
We use the CTEQ6L parton distributions \cite{Pumplin:2002vw} 
with 3 flavors of quarks and the next-to-leading-order formula 
for $\alpha_s(\mu)$ with 4 flavors of quarks and
$\Lambda_\textrm{QCD}$=326~MeV.
The factorization and renormalization scales
are set to $\mu=(Q^2+Q_T^2)^{1/2}$.
We impose a rapidity cut $|y|<1$ on the $J/\psi$ momentum.
 
\begin{figure}[t]
\begin{tabular}{lcr}
\epsfig{file=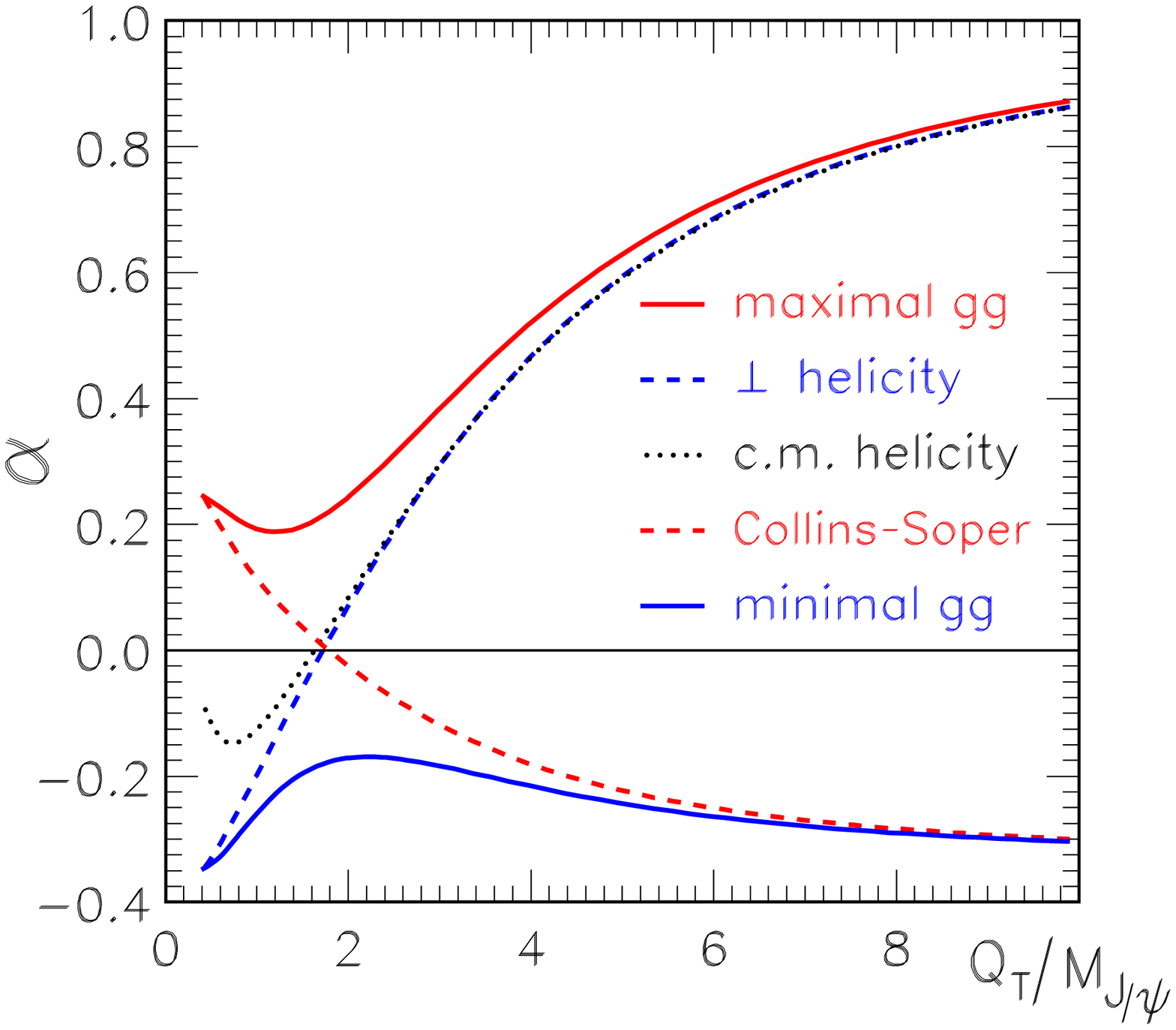,height=7cm}&\qquad&
\epsfig{file=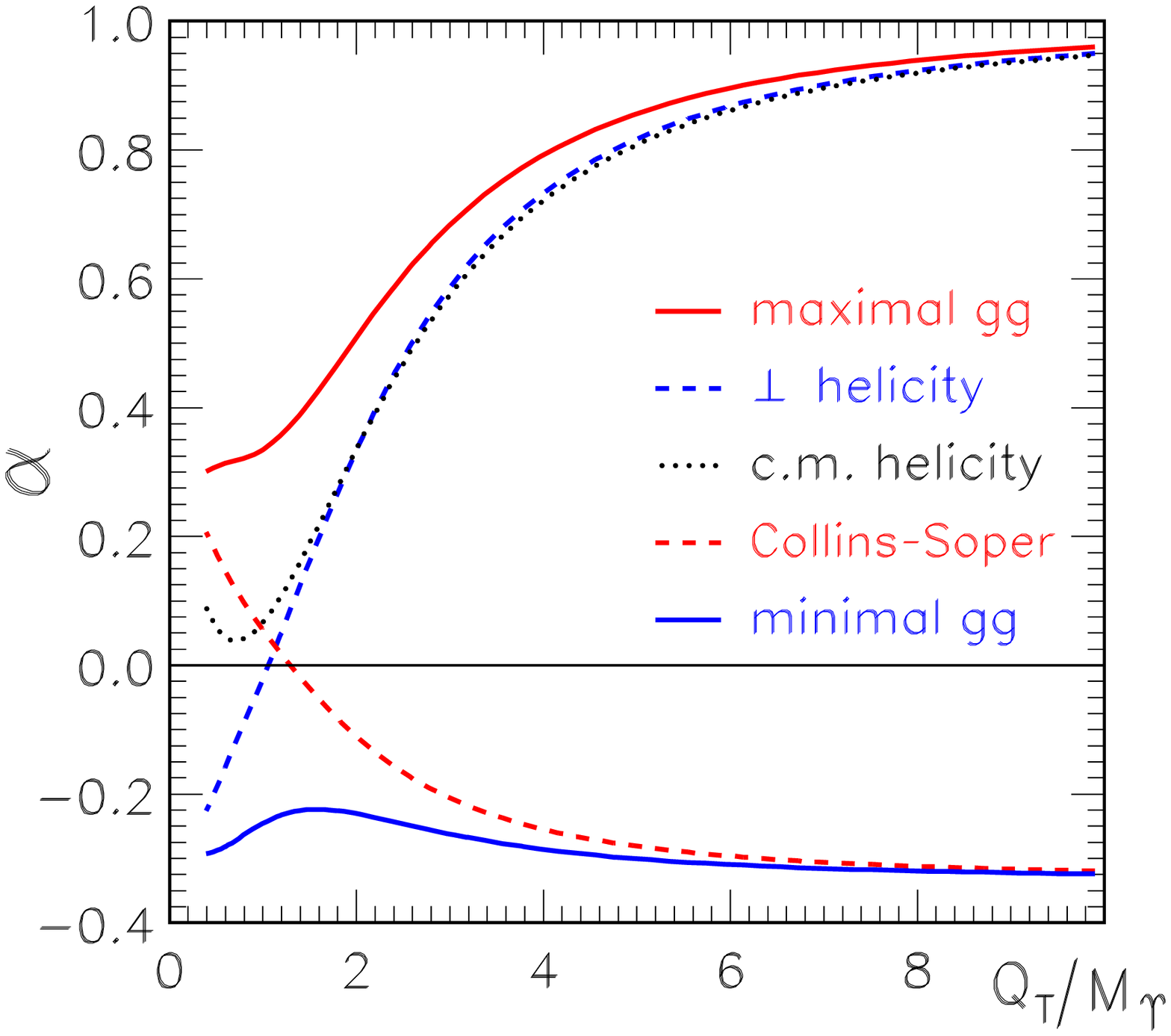,height=7cm}
\end{tabular}
\vspace*{-2ex}
\caption{  
Polarization variable $\alpha$ for various spin-quantization axes
as functions of $Q_T/M_{J/\psi}$ 
for direct $J/\psi$ at the Tevatron (left panel)
and direct $\Upsilon(1S)$ at the LHC (right panel).
The SQA's are the maximal and minimal $gg$ axes 
(upper and lower solid lines),
the perpendicular helicity and Collins-Soper axes 
(upper and lower dashed lines at large $Q_T$),
and the c.m.\ helicity axis (dotted line).
\vspace*{0ex}
}
\label{fig:alphaTev}
\end{figure}
%

A convenient polarization variable for $J/\psi$ is 
$\alpha = (\sigma_T - 2 \sigma_L)/(\sigma_T + 2 \sigma_L)$,
whose range is $-1 \le \alpha \le +1$.
We consider five SQA's: 
the maximal and minimal $gg$ axes defined by 
Eqs.~(\ref{ab:maxgg}) and (\ref{ab:mingg}) together with 
Eqs.~(\ref{deltacon}) and (\ref{ratio-opt}),
the perpendicular helicity and Collins-Soper axes 
defined by $X_{\perp {\rm h}}$ and $X_{\rm CS}$
in Eq.~(\ref{epsilonL-CS}),
and the {\it c.m.\ helicity axis}, which has been used 
in most previous work on this problem. 
The leading-order predictions for $\alpha$ for the five
SQA's are shown as functions of $Q_T/M_{J/\psi}$ in 
the left panel of Fig.~\ref{fig:alphaTev}.  
For the c.m.\ helicity and perpendicular helicity axes, 
$\alpha$ increases with $Q_T$ 
and asymptotically approaches $1$.  
The two axes are essentially identical at large transverse momentum.
For the maximal $gg$ axis, it approaches $1$ much more rapidly.
For the Collins-Soper axis, $\alpha$ decreases 
with $Q_T$, reaching $-0.33$ at $Q_T = 10~M_{J/\psi}$.
For the minimal $gg$ axis, $\alpha$ approaches 
the same asymptotic value much more rapidly. 

We also consider direct $\Upsilon(1S)$  production at the CERN LHC,
which is a $p p$ collider with c.m.\  energy 14~TeV.
For the NRQCD matrix elements, we use central CTEQ5L values from 
Ref.~\cite{Braaten:2000cm}:
$\langle O_1^{\Upsilon}(^3S_1) \rangle = 10.9~{\rm GeV}^3$, 
$\langle O_8^{\Upsilon}(^3S_1)\rangle = 0.025~{\rm GeV}^3$, 
$\langle O_8^{\Upsilon}(^1S_0)\rangle = 0.068~{\rm GeV}^3$, and
$\langle O_8^{\Upsilon}(^3P_0)\rangle/m_b^2 = 0.014~{\rm GeV}^3$.
We have used the averages of the values for the 
$\langle O_8^{\Upsilon}(^1S_0)\rangle = 0$ and
$\langle O_8^{\Upsilon}(^3P_0)\rangle = 0$ cases of Ref.~\cite{Braaten:2000cm}.
In the parton cross sections, we set $m_b = M_{\Upsilon}/2 = 4.7~{\rm GeV}$.
We imposed a rapidity cut $|y|<3$ on the $\Upsilon$.
The leading-order predictions for $\alpha$ for the five
SQA's are shown as functions of $Q_T/M_\Upsilon$ in 
the right panel of Fig.~\ref{fig:alphaTev}.

Our optimal SQA's are not useful for
fixed-target experiments, because the recoiling jet 
is usually not observed.  However, the angle $\theta_{\rm jet}$
of the recoiling jet can be measured relatively easily 
in high-energy hadron colliders. 
To provide some idea of how much the data sample will be decreased
by the requirement that $\theta_{\rm jet}$ be measured, 
we impose a cut on the pseudorapidity 
$\eta_{\rm jet} = \ln \tan(\theta_{\rm jet}/2)$ of the jet.  
For $J/\psi$ at the Tevatron,
the fraction of events satisfying the $J/\psi$ rapidity cut $|y| < 1$
that also survive a jet pseudorapidity cut $|\eta_{\rm jet}| < 1$
is greater than $0.21$ for $Q_T > M_{J/\psi}$.
For $\Upsilon(1S)$ at the LHC,
the fraction of events satisfying the $\Upsilon$ rapidity cut
that also survive a jet pseudorapidity cut $|\eta_{\rm jet}| < 3$
is greater than $0.72$ for $Q_T > M_\Upsilon$.
These fractions are large enough that measuring $\theta_{\rm jet}$ 
should not dramatically decrease the size of the data sample. 

The CDF and D0 Collaborations have measured the polarization 
as a function of $Q_T$ for charmonium and bottomonium mesons
produced at the Tevatron \cite{CDF:polznpsi,Acosta:2001gv,D0:2008za}.
For $J/\psi$, the variable $\alpha$ for the c.m.\ helicity axis
was measured for $Q_T/M_{J/\psi}$ as high as 9.7.
For $\Upsilon(1S)$, $\alpha$ for the c.m.\ helicity axis
was measured for $Q_T/M_\Upsilon$ as high as 2.1.
At the LHC, it should be possible 
to measure $\alpha$ for these and other heavy quarkonium 
mesons out to much larger values of $Q_T$.

The measurement of $\alpha$ for two different SQA's will provide 
more information about the spin if there is a large difference 
in the prediction of the transverse polarization 
with respect to the two axes.
Our leading-order calculations predict a large difference 
in $\alpha$ for the perpendicular helicity and Collins-Soper axes  
for $Q_T \gg M$.  One advantage of these two axes 
is that $X_{\rm CS}$  and $X_{\perp {\rm h}}$ in Eq.~(\ref{epsilonL-CS})
are invariant under independent longitudinal boosts 
of the two colliding hadrons.  This implies that $\alpha$ 
for these axes is insensitive to collinear radiation 
from the colliding partons, so radiative corrections  
may be smaller than for the c.m.\ helicity axis.
Our leading-order calculations predict an even larger
difference in $\alpha$ for the maximal and minimal $gg$ axes.
These optimal axes require the measurement of the direction
of a recoiling jet, but we have shown that this should not 
dramatically decrease the size of the data sample.

Similar methods could be used to derive optimal SQA's  
for the production of heavy elementary particles in the 
standard model, such as
the weak bosons $W^\pm$ and $Z^0$ and top quark.
There should be a large difference in the polarization between the 
perpendicular helicity and Collins-Soper axes,
but an even larger difference between 
the appropriate maximal and minimal SQA's.
If predictions for the polarization of these particles 
with respect to optimal SQA's can be verified,
then optimal SQA's will also provide a new window into 
the spins of new particles created at the LHC.

\begin{acknowledgments}
This research was supported in part by the Department of Energy 
under Grant Nos. DE-FG02-05ER15715 and DEFC02-07ER41457,
and by the KOSEF under Grant No. R01-2008-000-10378-0.
\end{acknowledgments}

{}

\end{document}